\documentclass[preprint,showpacs,preprintnumbers,amsmath,amssymb]{revtex4}

\usepackage{graphicx}
\usepackage{dcolumn}
\usepackage{bm}

\begin{document}


\title{Does The $\triangle I =1/2$ Rule Hold In D and B
$\rightarrow \pi \pi$ Decays? }

\author{Peng Guo}
\email{guopengnk@eyou.com}
\author{Xiao-Gang He}
\altaffiliation[On leave of absence from ]{Physics Department,
National Taiwan University} \email{hexg@phys.ntu.edu.tw}
\author{Xue-Qian Li}%
 \email{lixq@nankai.edu.cn}
\affiliation{%
Department of Physics, Nankai University, Tianjin, China\\
}%


\date{\today}

\begin{abstract}

{\small Although two pion decays of $K$, $D$ and $B$ have similar
isospin structures, there are dramatic differences in the ratios
$R_K$ and $R_{D,B}$ of amplitudes from $\Delta I =3/2$ and $\Delta
I =1/2$ interactions. In $K\to \pi\pi$ decays there is the famous
$\Delta I =1/2$ rule with $R_K \approx 1/22$, whereas in $B(D) \to
\pi\pi$ decays the ratios $R_{D,B}$ are of order one and therefore
there is no such a rule. In this work we study decay amplitudes in
$B(D)\to \pi\pi$ using  QCD factorization calculations paying
particular attention to discrepancies between data and theoretical
estimates. Since isospin does not play a special role in
factorization calculations, no $\Delta I =1/2$ rule is expected.
We find that theoretical results on the size of the amplitudes are
in qualitative agreement with data. However the phases for the
amplitudes are very different. We show that the effects of
re-scattering between the two pions in the final state can play a
crucial rule in understanding the differences between $B(D)\to
\pi\pi$ and $K\to \pi\pi$ decays. We also comment on the role of
isospin analysis which applies to the study of CP violation in
$B\to \pi \pi$ decays.}
\end{abstract}

\pacs{Valid PACS appear here}
\maketitle

It is well known for many years that there is a $\Delta I=1/2$
rule for $K\to \pi\pi$ decays. In these decays the ratio $R_K$ of
the isospin $I =2$ amplitude for the two final pions from $\Delta
I =3/2$ interaction to the $I =0$ amplitude from $\Delta I =1/2$
interaction is about 1/22 which is much smaller than naive
expectations. Why there is such a big difference between the two
amplitudes presents a serious challenge to theory. Many attempts
have been made to explain the $\Delta I =1/2$ rule. Since the
pioneer papers by Gaillard, Lee and Altarelli,
Maiani\cite{Gaillard}, it was hoped that the $\Delta I=1/2$ rule
may be explained by the running of the relevant Wilson
coefficients. However, the different evolution of the Wilson
coefficients of the operators which correspond to $\Delta I=1/2$
and $\Delta I=3/2$ respectively can only result in a factor of
$2\sim 4$, as the data demands a deviation of more than 20,
another factor of $6\sim 10$ is still missing. Other theoretical
scenarios have been proposed to explain the difference. For
example, the final state interaction \cite{Isgur}, the
contribution of the penguin-induced operator $O_6$ \cite{Shifman}
and even the diquark intermediate states \cite{Stech}.

Since the mesons $K$, $D$ and $B$ all have the same isospin
structure, it is natural to expect that in two pion decays of $B$
and $D$ there are similar $\Delta I = 1/2$ rules just like in
$K\to \pi\pi$ decays. It is interesting to study in more details
the isospin amplitudes of $B(D) \to \pi\pi$ decays. Moreover, one
hopes that the study of whether the $\Delta I=1/2$ rule holds or
not in B and D cases can provide crucial information to understand
the mechanism which enforces the $\Delta I =1/2$ rule in the K
case.

The decay amplitudes for $K$, $D$ and $B$ into two pions can be
parameterized in the same way as,
\begin{eqnarray}
&&A^{-0}(P^-\to \pi^-\pi^0) = \sqrt{{3\over 2}} A_2 e^{i\delta_2};\nonumber\\
&&A^{+-}(P^0\to\pi^+\pi^-) = {1\over \sqrt{3}} A_2 e^{i\delta_2} + \sqrt{{2\over 3}}
A_0 e^{i\delta_0},\nonumber\\
&&A^{00}(P^0 \to\pi^0\pi^0) = \sqrt{{2\over 3}} A_2 e^{i\delta_2}
- {1\over \sqrt{3}} A_0 e^{i\delta_0}, \label{iso}
\end{eqnarray}
where $P$ is one of the  $K^-,\;\;\bar K^0$, $D^-,\;\;D^0$ and
$B^-,\;\; \bar B^0$ mesons. $A_I$ and $\delta_I$ are the
amplitudes with the two pions in isospin ``I'' states and their
corresponding phases, respectively.

The amplitude $A_0$ is induced by an interaction with isospin 1/2,
whereas $A_2$ is induced by $ I = 3/2$ interaction. The $\Delta I
=1/2$ rule refers to the fact that the ratio $A_2/A_0$ is much
smaller than one. Using the above information one can write this
ratio in terms of branching ratios,
\begin{equation}\label{RP}
R_{P}=|{A^{(P)}_2\over A^{(P)}_0}|=[{\Gamma_P^{+0}\over {3\over
2}(\Gamma_P^{+-}+\Gamma_P^{00})-\Gamma_P^{+0}}]^{1/2}.
\end{equation}

\vspace{0.5cm}
\begin{table}[htb]
\caption{Branching ratios. $D\to \pi\pi$ and $B\to \pi\pi$ are
from Ref.\cite{Data} and Ref.\cite{Belle}, respectively.}
\vspace{0.5cm}
\begin{tabular}{|l|c|c|c|c||}
\hline
  & Data  & naive factorization & QCD factorization  \\
  \hline
  \hline
  BR($D^{+} \rightarrow \pi^{+} \pi^{0}$) & $(2.6 \pm
0.7) \times 10^{-3} $  &
$2.97 \times 10^{-3} $ & $2.88  \times 10^{-3} $ \\
\hline BR($D^{0} \rightarrow \pi^{+} \pi^{-}$) & $(1.38 \pm 0.05)
\times 10^{-3}$
& $2.02  \times 10^{-3} $ & $2.08  \times 10^{-3} $\\
\hline BR($D^{0} \rightarrow \pi^{0} \pi^{0}$) & $(8.4 \pm 2.2)
\times 10^{-4}$ & $5.51  \times 10^{-6} $ &  $3.65  \times 10^{-6}
$
\\ \hline
 BR($D^{0} \rightarrow \pi^+ \pi^- +\pi^0\pi^0$) & $(2.22 \pm 0.23) \times
10^{-3}$ &
$2.02  \times 10^{-3} $ &  $2.08  \times 10^{-3} $ \\
\hline $R_D$ & $0.65\pm 0.13$&0.79&0.75\\
\hline\hline
 BR($B^{+} \rightarrow \pi^{+} \pi^{0}$) & $(5.5 \pm
0.6) \times 10^{-6} $  &
$6.59  \times 10^{-6} $ &  $5.84  \times 10^{-6} $\\
\hline BR($B^{0} \rightarrow \pi^{+} \pi^{-}$) &  $(4.6 \pm 0.4)
\times 10^{-6} $ &
$ 8.23  \times 10^{-6} $ & $8.38  \times 10^{-6} $ \\
\hline BR($B^{0} \rightarrow \pi^{0} \pi^{0}$) &  $(1.51 \pm 0.6)
\times 10^{-6} $ & $2.29 \times 10^{-7} $ &  $2.48  \times 10^{-7} $\\
\hline BR($B^{0} \rightarrow \pi^+ \pi^-+\pi^0\pi^0$) &  $(6.11
\pm 0.52) \times 10^{-6} $ &
$ 8.46  \times 10^{-6} $ &  $8.63  \times 10^{-6} $  \\
\hline $R_B$& $1.11\pm 0.20$&0.96& 0.84\\
\hline
\end{tabular}
\label{table1}
\end{table}

The branching ratios for $D\to \pi\pi$ have been measured to good
precision\cite{Data}. Recently the $B\to \pi\pi$ branching ratios
have also been measured\cite{Belle}. In Table \ref{table1} we list
the branching ratios of the $\pi\pi$ decays of $D$ and $B$. With
these measured branching ratios, one can easily check if there are
$\Delta I =1/2$ rule in $B$ and $D$ to $\pi\pi$ decays. We obtain

\begin{eqnarray}
R_D=0.67\pm 0.13;\;\;\;\;R_B = 1.11\pm 0.20.
\end{eqnarray}
It is obvious that the $\Delta I =1/2$ rule is violated in $B
(D)\to \pi\pi$ decays. The situation is very different from that
in $K \to \pi\pi$ decays. It is therefore important if one can
understand the situation by studying these decays more carefully.

To have a full understanding of the situation, one also needs to
consider phase shifts. From eq. (\ref{iso}) one obtains the
isospin phase differences $\cos\delta_P =\cos(\delta_0
-\delta_2)_P$ as

\begin{eqnarray}
\cos\delta_P = {3|A^{+-}|^2-6|A^{00}|^2 + 2|A^{-0}|^2\over
4\sqrt{3} |A^{-0}|\sqrt{|A^{+-}|^2+|A^{00}|^2-2|A^{-0}|^2/3}}.
\label{fsi1}
\end{eqnarray}

Using experimental data, we obtain
\begin{eqnarray}
&&\cos\delta_D=0.13\pm 0.06,\;\;\mbox{for}\;\; D\to \pi\pi,\nonumber\\
&&\cos\delta_B =0.58\pm 0.20, \;\;\mbox{for}\;\; B\to \pi\pi.
\label{fsi}
\end{eqnarray}
These results indicate that the difference of the phase shifts
$\delta_P$ is sizeable. In the above we have neglected CP
violation in the decay amplitude which is applicable for $D\to
\pi\pi$ decays. If CP violation is sizeable which may happen in
$B\to \pi\pi$ decays, one needs to replace $\cos\delta_P$ by
$\cos(\delta_P + \phi_P)$ with $\phi_P$ being the CP violating
phase difference in the decay amplitudes. We will come back to
this later.

We first discuss the quantity $R_P$. In the past few years
progresses have been made in the calculations of a heavy meson
decays into two light mesons based on QCD factorization. Several
ways of calculating $B\to \pi\pi$ have been developed with
different methods treating non-purterbative quantities
involved\cite{Beneke,other,li,prijol}. There are also many
model-independent studies about the amplitudes for $B\to \pi\pi$
decays based on isospin and/or SU(3) symmetries\cite{fsi,new}. We
will take QCD improved factorization as the theory to compare with
data. We obtain the isospin amplitudes in the following

\begin{eqnarray}
A_0 &=& -\{\sqrt{{2\over 3}} \lambda^{'}_{u}
(a_{1}-\frac{a_{2}}{2})+ \sqrt{{3\over 2}}
\lambda^{'}_{p}[a^{p}_{4}+\frac{a_{7}}{2}-\frac{a_{9}}{2}+\frac{a^{p}_{10}}{2}
+ \gamma^{\pi}_{X}(a^{p}_{6}+\frac{a^{p}_{8}}{2})]\}A_{\pi\pi}
\nonumber  \\
 &&-\sqrt{{3\over 2}}[\lambda^{'}_{u}
b_{1}+(\lambda^{'}_{u}+\lambda^{'}_{c})(b_{3}+2b_{4}-\frac{b^{EW}_{3}}{2}
+\frac{b^{EW}_{4}}{2})]B_{\pi \pi},  \nonumber \\
 A_2&=&-{1\over
\sqrt{3}}[\lambda^{'}_{u}(a_{1}+a_{2})+\frac{3}{2}\lambda^{'}_{p}
(-a_{7}+\gamma^{\pi}_{X}a^{p}_{8}+a_{9}+a_{10}^{p})]A_{\pi\pi}.
\label{fact}
\end{eqnarray}
In the above expressions, $\lambda'_{q} = V_{qb}V^*_{qd}$, $A_{\pi
\pi}=i (G_{F}/\sqrt{2})(m^{2}_{B}-m^{2}_{\pi})F^{B \rightarrow
\pi}_{+}(0)f_{\pi}$, and $\gamma^{\pi}_{X}(\mu)=2m^{2}_{\pi}/
\overline{m}_{b}(\mu) (
\overline{m}_{u}(\mu)+\overline{m}_{d}(\mu) )$. The other
quantities are defined in Ref.\cite{Beneke}

Terms proportional to $b_i$ in eq.(\ref{fact}) are referred as
annihilation contributions. These terms have end-point divergences
of the form $X_A = \int^1_0 \phi_\sigma (y) dy/(1-y)$ and need
regularization. The term proportional to $f^{II}$ also has a
similar end-point divergence $X_H$. These divergences associated
with $1/m_{b,c}$ corrections indicate in a way the incompleteness
of factorization calculation. In Ref.\cite{Beneke} these end-point
divergences are parameterized as $\ln(m_b/\Lambda_h) (1+\rho_{A,H}
e^{i\phi_{A,H}})$. An idea of the size of the corrections can be
obtained by varying $\rho_{A,H}$ and $\phi_{A,H}$.

To finally obtain numerical results, one needs to know the CKM and
the hadronic parameters in the above amplitudes. There are
considerable progresses in the determination of the CKM
parameters\cite{Data,ciuchini}. We will use the central values
given in Ref.\cite{Data} with $s_{12} = 0.2243$, $s_{23} =
0.0413$, $s_{13} = 0.0037$ and the CP violating phase
$\gamma(\delta_{13}) = 60^\circ$ for illustration. For the
hadronic parameters we use the ``default'' values given in
Ref.\cite{Beneke}. The results for the ratio $R_{B}$ are listed in
Table \ref{table1}. In the table, we also list the values obtained
by using naive factorization\cite{ali} for comparison. We see that
the $\alpha_s$ order correction is at the order of 10\% and can be
as large as 30\% if the involved hadronic parameters vary within a
reasonable range. One also notes that the term $f_I$ generates an
absorptive part which is absent in naive factorization
calculations.

One can see from Table \ref{table1} that the theoretical
calculation is in qualitative agreement with the data that there
is no $\Delta I = 1/2$ rule in $B\to \pi\pi$ decays. This is
expected since that in factorization calculations the isospin does
not play a special role. The leading operators, i.e. the tree
operators, which contribute to $B \to \pi\pi$, contain both
$\Delta I=1/2$ and $ 3/2$ pieces with similar weights, the
amplitudes for isospin I =0 and I =2 are therefore expected to
have similar sizes.

We now consider the situation for $D\to \pi\pi$ decays. In Ref.
\cite{Sannino} Sannino noticed the violation of the $\Delta I=1/2$
rule in $D\rightarrow \pi\pi$ and tried to understand it based on
the effective weak Hamiltonian. Assuming that the hadronic matrix
elements $M_2$, $M_0$ and $\tilde{M_0}$ defined in
Ref.\cite{Sannino} are all equal, numerically $R_D\sim 0.32 \sim
0.44$ which is not too far away from data.

The c-quark is not as heavy as the b-quark, factorization
calculation for $D\to \pi\pi$ may not work as well as for $B\to
\pi\pi$. We however expect that a theoretical calculation based on
QCD factorization scheme can still provide some crude estimate. We
therefore also use the QCD factorization for $D\to \pi\pi$. In
this calculation we will use the Wilson coefficients given in
Ref.\cite{Buccella}. We have also checked the dependence on the
renormalization scale and find that the result on $R_D$ does not
change much with respect to the scales. The numerical results are
listed in Table \ref{table1}. We see that the situation is similar
to that for $B\to \pi\pi$, there is no $\Delta I =1/2$ rule.

The calculation for $K \to \pi\pi$ using factorization  becomes
very questionable. Nevertheless, attempts have been made to
evaluate the amplitudes.
Buras et al.\cite{Buras} evaluated the ratio $R_K$ based on the
large N expansion approach with
\begin{equation}
R_K=\mid \frac{A_{0}}{A_{2}} \mid
=\frac{c_{1}(\mu)<O_{1}(\mu)>_{0}+c_{2}(\mu)<O_{2}(\mu)>_{0}+
\sum^{6}_{i=3}
c_{i}(\mu)<O_{i}(\mu)>_{0}}{c_{1}(\mu)<O_{1}(\mu)>_{2}+c_{2}(\mu)<O_{2}(\mu)>_{2}}
\end{equation}
and the hadronic matrix elements can be parameterized as
\begin{eqnarray}
&&<O_{1}>_{0}=-\frac{1}{9}X B_{1}^{(1/2)},\;\;
<O_{2}>_{0}=\frac{5}{9}X
B_{2}^{(1/2)}, \nonumber\\
&&<O_{3}>_{0}=\frac{1}{3}X B_{3}^{(1/2)},\;\;
<O_{4}>_{0}=<O_{3}>_{0}+<O_{2}>_{0}-<O_{1}>_{0},\nonumber\\
&&<O_{5}>_{0}=\frac{1}{3}<O_{6}>_{0},\;\;
<O_{1}>_{2}=<O_{2}>_{2}=\frac{4\sqrt{2}}{9}X
B_{1}^{(3/2)},\nonumber\\
&& <O_{6}>_{0}=-4\sqrt{\frac{3}{2}}
[\frac{m_{K}^{2}}{\overline{m}_{s}(\mu)+\overline{m}_{d}(\mu)}]^{2}
\frac{F_{\pi}}{\kappa}B_{6}^{(1/2)},
\end{eqnarray}
where $X=\sqrt{\frac{3}{2}}F_{\pi}(m_{K}^{2}-m_{\pi}^{2})$,
$\kappa=F_{\pi}/(F_{K}-F_{\pi})$ and $B_{i}$ are the hadronic
parameters to be determined either by fitting experimental data or
invoking phenomenological models. By fitting data they obtained a
set of the hadronic parameters as $B_1^{(3/2)}=0.48$,
$B_1^{(1/2)}=10$, $B_2^{(1/2)}=5$ and $B_3^{(1/2)}=B_6^{(1/2)}=1$.
The value of $R_K$ calculated in this approach agrees with data.
One notes that several bag parameters $B_i$ are substantially away
from the naive value of $B_{i}=1$. Thus one expects that some
non-perturbative effects and the final state interaction effects
are involved altogether in the parameters.

We note that although theoretical calculations agree with data,
one can be confirmed with the fact that there is not a $\Delta I
=1/2$ rule in $B (D) \to \pi\pi$ decays, there is a large
difference in $\bar B^0 (D^0) \to \pi^0\pi^0$ branching ratio
between naive theoretical estimates and data. It is important to
see if these discrepancies can be explained.

A possible source may be due to uncertainties in the factorization
calculation, in particular corrections of $1/m_{b,c}$. As pointed
out earlier that the end-point divergences appearing in the hard
scattering and annihilation signal incompleteness of the
$1/m_{b,c}$ corrections. A complete treatment of these effects are
beyond the scope of this work. To have some idea about the effects
of $1/m_{b,c}$ we have carried out a calculation by varying the
parameters $\rho_{AH}$ from 0 to 3 and $\phi_{A,H}$ from 0 to
$2\pi$. We find that the changes on the branching ratios of
$D^+(B^+) \to \pi^+\pi^0$ and $D^0(B^0) \to \pi^+\pi^-$ are in the
range of 20\% to 30\%, and the changes on $D^0 (B^0)\to
\pi^0\pi^0$ can be dramatic (a factor of 2 to 3). However, it is
still not possible to bring $D^0(B^0)\to \pi^0\pi^0$ branching
ratios to their experimental values. One may need to consider
other effects which are not included in the factorization
approximation.

To this end we consider long distant final state interaction
effects. If the differences are really due to FSI effects, it
should not only explain the differences mentioned above, but
should also explain why there is a $\Delta I =1/2$ rule in $K\to
\pi\pi$, but not in $B(D)\to \pi\pi$.

To explain the $\Delta I=1/2$ rule in K-decays, Isgur et al.
\cite{Isgur} proposed a possible mechanism for the smallness of
$R_K$ ($\sim 1/22$) that before the two pions produced from the
weak decay of the kaon fly apart to become free particles, they
reside in  bound states of either $I=0$ or $I=2$. Due to different
interactions for the two different isospin states, the
wavefunctions at origin for $I=0$ and $I=2$ states are distorted
differently. The hadronic matrix elements of the two isospin
states would undergo an enhancement or a suppression
as\cite{Isgur}
\begin{eqnarray}
\frac{<(\pi \pi)_{I=0} \mid H^{1/2}_{W} \mid K>}{<(\pi \pi)_{I=2}
\mid H^{3/2}_{W} \mid K>} =(\frac{d_{0}}{d_{2}})^{\frac{1}{2}}
\frac{<(\pi \pi)_{I=0}^{free} \mid H^{1/2}_{W} \mid K>}{<(\pi
\pi)_{I=2}^{free} \mid H^{3/2}_{W} \mid K>}.\label{isgur13}
\end{eqnarray}
Here the distorting parameter $d_I$ is defined as $d_I =
|{\psi^{true}(0)/\psi^{free}(0)}|^2$.

The general requirement for the interaction potential is that the
$I=0$ channel experiences an attractive interaction whereas the
$I=2$ channel experiences a repulsive one. We note that a simple
square well potential of the form $U {\bf I_1}\cdot {\bf I_2} =
(1/2)U[I(I+1) - I_1(I_1+1)-I_2(I_2+1)]$ with $U>0$ can result in
the required potential form.

Fitting data on the low-energy $\pi-\pi$ scattering phase shifts,
the parameters of $V_I$ and $a_I$ can be determined. The
distortion due the potential can result in an enhancement of the
channel of the $\Delta I=1/2$ over the $\Delta I=3/2$ by a factor
$r=(d_{0}/d_{2})^{1/2}$ for $K\rightarrow \pi\pi$ which can be as
large as 9 $\sim$ 10. For the isospin correlated potential $U {\bf
I_1}\cdot {\bf I_2}$ with $U = 0.4$ GeV and potential range
$a_0=a_2 = 0.8$ fm, the enhancement factor $r$ is 9. This factor
can make up the gap between theory and experimental data leading
to the $\Delta I=1/2$ rule in the case of $K \to \pi\pi$. One can
also try more complicated potential forms, such as Gaussian-type
potentials as studied in Ref.\cite{Isgur}, but the qualitative
features are unchanged.

Since the final two pions produced from $B$ and $D$ decays have
similar isospin structure except that their kinetic energies are
much higher, they should experience similar effects due to the
potential. However, since the two pions have much higher kinetic
energies they have much shorter time to interact with each other
before emerging out of the bound states to produce large effects.
More specifically, in the D and B cases, the momenta $|{\bf
k}|={1\over 2}\sqrt{(M_{D(B)}^2-M_{\pi}^2)}$ are much larger than
the potential energies $|V_I|$, one can easily verify that
$d_0\approx d_2\approx 1$. Even though for the $B$ and $D$ decays,
the two pions fly very fast and the non-relativistic quantum
mechanical scenario may not produce accurate numbers, the
qualitative conclusion about smallness of the bound state effects,
however, should hold. We therefore see that this mechanism can
easily explain why there is a $\Delta I = 1/2$ rule for $K \to
\pi\pi$ but not for $B(D) \to \pi\pi$.

We now discuss the FSI phases. Using experimental data, we obtain
$\cos\delta_D$ and $\cos\delta_D$ which are shown in eq.(5). These
results indicate that the difference of the phase shifts
$\delta_P$ is sizeable. Although factorization amplitudes in
eq.(\ref{fact}) have included some re-scattering effects at quark
level, they are not large enough to explain data. Factorization
calculations using the default values for relevant hadronic
parameters would give a $\cos\delta_P$ to be very close to 1.
There are uncertainties in factorization calculations in
regularizing the end-point divergences. As mentioned earlier, we
have checked numerically that within reasonably allowed parameter
space, these contributions cannot generate large enough phase
shifts to reproduce data.

The distorting effects in eq.(\ref{isgur13}) discussed earlier can
also generate a phase shift. At the kaon energy, it is possible to
generate the required phase shift for $K \to \pi\pi$. However
these phase shifts go to zero as the energies become much larger
than the potentials. For $B$ and $D$ decays, the phase shifts are
practically zero because the pions from $D(B)$ decays have high
energy and pass through the potential in too short a time to
produce any significant phase shifts. There is a need of
additional effects. The effects producing large phase shifts in
$D$ and $B$ decays must be different from the mechanism in
Ref.\cite{Isgur} which is indeed available, so that at higher
energies more channels become active in producing absorptive part
of decay amplitudes and therefore phase shifts\cite{fsi,new}.

From the calculated branching ratios of $D^0\rightarrow \pi\pi$
and $B^0\rightarrow \pi\pi$ in Table \ref{table1}, we note that
the value for $\pi^+\pi^-$ is obviously higher than data, whereas
the value for $\pi^0\pi^0$ is significantly lower than data.
However the sum of $BR(P^0 \to \pi^+\pi^-)$ and $BR(P^0\to
\pi^0\pi^0)$ , and also $BR(P^- \to \pi^-\pi^0)$ are close to
experimental values in both $D$ and $B$ cases. Since only these
sums determine the ratios $R_P$ (see eq.(2)), the theoretical
calculations are close to data. One should keep in mind that there
are uncertainties in many of the hadronic parameters. The
theoretical predictions for the branching ratios can change, but
it is difficult to generate large enough branching ratios for $B^0
(D^0)\to \pi^0\pi^0$. If re-scattering of the two pions can change
the higher valued $\pi^+\pi^-$ into the lower valued $\pi^0\pi^0$
after the pions come out of the bound states as discussed above,
it may be able to produce the correct values to meet data. In fact
large phase shifts due to re-scattering in $B(D)$ decay into two
light mesons have been noticed before\cite{fsi,new}. We have
carried out a simple exercise similar to some work in
Ref.\cite{fsi} taking the magnitudes of $A_{0,2}$ as determined by
factorization calculations and $\delta$ as a free parameter to fit
data. We find that with $\delta_D$ equal to $0.45\pi$,
$\cos\delta_D$ is close to the central value of the data. In the
case of $B\to \pi\pi$, with the default values for the hadronic
parameters, $\delta_B = 0.30\pi$ can reproduce a $\cos\delta_B$
which makes the branching ratio of $B\rightarrow \pi^0\pi^0$ to be
close to the experimental central value. In this case both
amplitudes $A_{0,2}$ are slightly higher than data, but can easily
be made to agree with data if a slightly smaller form factor
$F^{B\to \pi} = 0.23$ is used instead of the default value of
$0.28$. A more precise theoretical evaluation of the re-scattering
phases would be difficult because the mechanisms are not fully
understood yet.

We finally make some comments about isospin analysis and CP
violation. The FSI phases are very important for the study of CP
violations. Sizeable CP asymmetry can occur in $B\to \pi\pi$
decays. We therefore will concentrate on $B\to \pi\pi$ decays. The
isospin amplitude given before can be further decomposed into
different components according to the CKM matrix elements
associated with

\begin{eqnarray}
A_i = V_{ub}V_{ud}^*a^T_i + V_{tb}V_{td}^*a^P_i.
\end{eqnarray}
The components $a^T_i$ mainly come from tree amplitudes by
exchange W-boson but also small corrections from loop-induced c
and u penguin contributions, and $a^P_i$ are induced at loop level
by c and t penguins.

In general there can be 4 complex hadronic parameters $a^T_{1,2}$
and $a^P_{0,2}$ to describe $B\to \pi\pi$ decays. Among them one
can always set one of the components to be real, there are
actually only 7 independent parameters. Furthermore in the SM, the
leading order from c and t penguin contributions to $a^P_2$ are
dominated by t penguin from operator $O_{9,10}$ which have the
same Lorentz structure as the tree operators. One has\cite{rosner}
$a^P_2 = {3/2}(c_9+c_{10})/(c_1 +c_2)a^T_2$. Neglecting other
smaller contributions to $a^{T,P}_2$ components, one only needs
five hadronic parameters to describe $B\to \pi\pi$ decays. This is
an important fact which has interesting implications.

One of them is that it can be used to determine CP violating phase
$\gamma$ if the 7 possible experimental observables in $B\to
\pi\pi$ decays are all measured (the 3 branching ratios, the 2
direct CP violation ($B^-\to \pi^-\pi^0$ has very small CP
violation in the SM and probably cannot be measured), and 2 mixing
induced CP violation for $\bar B^0\to \pi^+\pi^-, \pi^-\pi^0$),
the 7 observables can determine the five hadronic parameters and
two CKM parameters $\rho$ and $\eta$. We need to wait for more
data to carry out a full analysis. If one takes the CKM parameters
as determined by other data, the hadronic parameters can already
be extracted.

In obtaining the value $\cos\delta_B$ in eq.(\ref{fsi}), we have
neglected possible CP violating effect in the isospin amplitudes.
We therefore cannot determine $a^T_0$ and $a^P_0$ separately. One
needs more data points to obtain more detailed information about
the size of the decay amplitudes and also phases. Fortunately
experimentally, besides the branching ratios listed in Table
\ref{table1} for $B\to \pi\pi$, there are also some measurements
on the direct CP violating parameter $A_{CP}$ and mixing induced
CP violating parameter $S_{CP}$\cite{Belle},
\begin{eqnarray}
&&A_{CP}(\bar B^0 \to \pi^+ \pi^-) = 0.37\pm 0.11,\;\;S_{CP}(\bar
B^0\to \pi^+\pi^-) = -0.61\pm 0.14\nonumber\\
&&A_{CP}(\bar B^0\to \pi^0\pi^0) = 0.28\pm 0.39.
\end{eqnarray}

Using five data points, the three branching ratios and the CP
asymmetry parameters $A_{CP}$ and $S_{CP}$ for $\bar B^0 \to
\pi^+\pi^-$, as input we can solve for the hadronic amplitudes. We
have

\begin{eqnarray}
1)\;\;\;\;&&a^T_2 = 0.366 A,\;\; a^T_0 = 0.276
e^{-i48.5^\circ}A,\;\;a^P_0=0.067e^{-i71.1^\circ}A;\nonumber\\
2)\;\;\;\;&&a^T_2 = 0.366A,\;\;a^T_0=0.289
e^{i54.3^\circ}A,\;\;a^P_0=0.065 e^{-i16.6^\circ}A.
\end{eqnarray}
Here $A=\sqrt{BR(B^-\rightarrow \pi^-\pi^0)\Gamma_{B_{total}}}$
Factorization calculations give different values, in particular
the phases. Improved theoretical framework of calculating the
amplitudes is needed.

If the penguin amplitude $A_0^P$ is neglected, there is no direct
CP violation, one can use eq. (\ref{fsi1}) to determine the FSI
phase. Since $A_0^P$ is small compared with the three amplitudes,
the use of eq. (\ref{fsi1}) is a good approximation. We note that
the two solutions have different FSI phases for each individual
amplitude, but the phase difference of $A^T_0$ and $A^T_2$ are
similar in size for solutions 1) and 2) and different in signs,
Since the cosine function is not sensitive to the sign, one would
obtain similar $\cos\delta_B$. One needs to find some observables
which can distinguish these solutions. We indeed find that CP
violation in $A_{CP}$ of $\bar B^0 \to \pi^0\pi^0$ depends on the
sign. We have not used the data as input because the error is
large there. Using the above amplitudes and phase we can predict
its value. We obtain $A_{CP}(\pi^0\pi^0) = -0.60$ and $0.18$ for
solutions 1) and 2), respectively. The solution 2) obtains a value
close to the current central value. But due to the large error at
present, it is too
 early to decide which solution is the correct one. When more
precise data become
available one can distinguish these solutions and obtain more
detailed information about the isospin amplitudes.

In summary in this work we have analyzed the isospin amplitudes
due to $\Delta I=1/2$ and $\Delta I =3/2$ interactions in the
$K\rightarrow \pi\pi$, $D\rightarrow \pi\pi$ and $B\rightarrow
\pi\pi$ decays. Experimental data clearly show that unlike the
situation for $K\to \pi\pi$, there is no $\Delta I = 1/2$ rule in
$B(D) \to \pi\pi$ decays. Theoretical calculations using the naive
factorization and the QCD factorization are consistent with data
about the violation of $\Delta I=1/2$ rule, but there are
difficulties to obtain correct phases in the amplitudes and
branching ratio for $BR(\bar B^0\rightarrow \pi^0\pi^0)$.

The question  why there is a $\Delta I =1/2$ rule in $K\to \pi\pi$
whereas not for $B(D) \to \pi\pi$ can be interpreted by the
re-scattering of the final two pions due to a isospin correlated
interaction proposed by Isgur et al. We note that the potential of
the form $U {\bf I_1}\cdot {\bf I_2} = (1/2)U[I(I+1) -
I_1(I_1+1)-I_2(I_2+1)]$ can play such a role. To fully explain
$B(D) \to \pi\pi$ data, there is also the need of re-scattering
after the pions come out of the bound states. These additional
phases are not precisely calculable at present.

We have also shown that isospin analysis for $B\to \pi\pi$
provides valuable information about CP violation in B decays and
hadronic amplitudes.

\noindent{Acknowledgement}

This work is partially supported by the National Natural Science
Foundation of China. We thank T. Barnes for explaining some
details about Ref.\cite{Isgur}.

\end{document}